\documentclass[11pt]{article}
\usepackage{moriond,epsfig}

\def\NIMA{{\em Nucl. Instrum. Methods} A}
\def\NPB{{\em Nucl. Phys.} B}
\def\PLB{{\em Phys. Lett.}  B}

\def\PRD{{\em Phys. Rev.} D}

\def\be{\begin{equation}}
\def\ee{\end{equation}}
\def\bea{\begin{eqnarray}}
\def\eea{\end{eqnarray}}

\def\numtonut{{\mathrm\nu}_{\mathrm\mu}\rightarrow{\mathrm\nu}_{\mathrm\tau}}
\def\nuetonut{{\mathrm\nu}_{\mathrm e}\rightarrow{\mathrm\nu}_{\mathrm\tau}}

\def\nut{{\mathrm\nu}_{\mathrm\tau}}
\def\stt{\sin^22\theta_{\mu\tau}}
\def\stte{\sin^22\theta_{e\tau}}
\def\tauu{{\mathrm\tau}}

\def\num{{\mathrm\nu}_{\mathrm\mu}}

\def\numb{\overline{{\mathrm\nu}}_{\mathrm\mu}}
\def\dmsq{\Delta\mathrm{m}^2_{\mathrm\mu \tau}}
\def\dmsqe{\Delta\mathrm{m}^2_{\mathrm e \tau}}

\begin{document}
\vspace*{4cm}
\title{Results from the short baseline experiments at CERN}

\author{Alfredo G.\ Cocco}

\address{Istituto Nazionale di Fisica Nucleare - Sezione di Napoli. \\
         Complesso Universitario di Monte S.Angelo - Napoli, I-80126 Italy}

\maketitle

\abstracts{Considerations based on particle physics (in particular the so-called
see-saw mechanism) and on cosmology, in relation to the Dark Matter question, 
motivated the search for $\numtonut$ oscillation in the region of small mixing
angles. We present here the outcome of the short-baseline programme
carried out at CERN by the CHORUS and NOMAD experiments. Both experiments are about 
to publish their final results. At values of $\dmsq$ greater than 1 eV$^2$ oscillations
between $\num$ and $\nut$ are excluded at 90\% C.L. down to  
$\stt \sim O(10^{-4}) $.}

\section{Introduction}

The discovery of neutrino oscillations represents today the most powerful
and elegant way
to establish the non zero mass of the neutrino and a non vanishing mixing
matrix between weak eigenstates in the leptonic sector~\cite{pontec}.

Among neutrino oscillation searches the $\numtonut$ channel is a highly
promising one to pursue with accelerators; 
given the high mass of the $\tau$ lepton it is
in fact relatively easy to produce a neutrino beam in which $\nut$ are
practically absent. The detection of $\nut$ charged-current
interactions in such a beam will prove with no doubt the mixing among 
different neutrino flavors.

Indications on the possible presence of mixing in the $\mu - \tau$
sector at small mixing angles and at cosmological relevant squared mass 
difference in relation to the question of Dark Matter~\cite{harari},
led to the construction of CHORUS and NOMAD, two short-baseline experiments
using different techniques to detect $\nut$ appearance.
So far, there is no evidence from these experiments of such a phenomenon 
even though the data analysis is not yet completed.

\section{Short-baseline programme at CERN}

Two short-baseline experiments have recently taken data at CERN and are about to
finalize their data analysis. Both of them aim at the detection of
$\numtonut$ oscillation via the $\nut$ appearance method, using the CERN SPS Wide
Band Neutrino Beam~\cite{spswbb}.
The beam contains predominantly muon neutrinos from $\pi^+$ and $K^+$ decay, 
with contamination levels for $\numb$ and $\nu_e, \overline{\nu}_e$ of, 
respectively, 5\% and 1\%. 
Neutrinos are mostly produced in a 290 m long decay tunnel, at an average
distance of about 625 m from the experimental hall. 
The estimated $\nut$ background is of 
the order of $3.3 \times 10^{-6}$ $\nut$ charged-current interactions per 
$\num$ charged-current interaction\cite{BDV} and is therefore negligible. 
The $\num$ component of the beam has an average energy of $27$ GeV.

\subsection{The CHORUS experiment}

\begin{figure}[t]
\begin{center}
\epsfxsize=25pc
\epsfbox{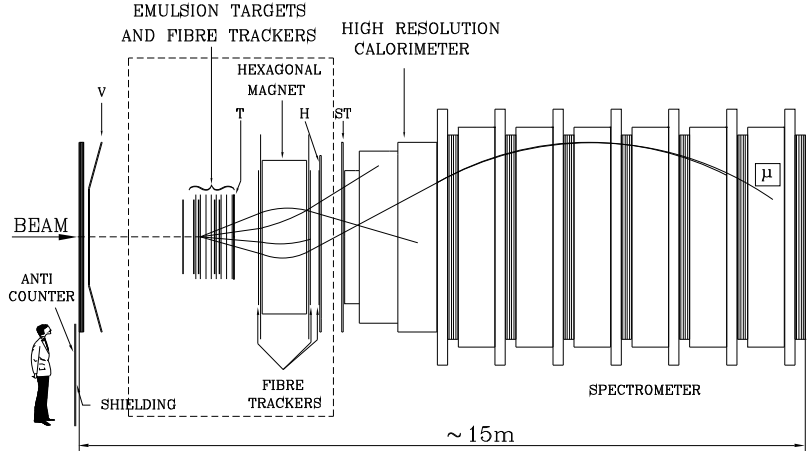}
\end{center}
\caption{\label{detector} General layout of the CHORUS detector.}
\end{figure}

The CHORUS apparatus~\cite{chdet}
is shown in Figure~\ref{detector}; it comprises an emulsion target, a
scintillating-fiber tracker system, trigger hodoscopes, a magnetic spectrometer, 
a lead scintillating-fiber calorimeter and a magnetized iron muon spectrometer.

The emulsion target has the unprecedented mass of $770 \; \mathrm{kg}$ and a surface
area of $1.42 \times 1.44 \; \mathrm{m}^2$. 
Neutrino interactions occur in nuclear emulsion, 
whose exceptional spatial resolution (below 1 $\mathrm{\mu m}$) 
and hit density ($300$ grains/$\mathrm{mm}$ along the track) allow
a complete three-dimensional reconstruction of the event.
The direct evidence of $\tauu$ production and decay is thus obtained analyzing
the trajectories of the charged particles near the interaction vertex.
A ``kink'' characterizes interesting events.

Electronic detectors downstream the emulsion target are used to locate the event
and, to a lesser extent, to reconstruct the event kinematics.

The detector has been exposed to the neutrino beam from 1994 to 1997, for a total 
of $5.06\times10^{19}$ protons on target.
A total of about $1\times10^{6}$ events have been reconstructed in the electronic
detectors to have the interaction vertex in the emulsion.

The number of neutrino interactions located in the emulsion target until now is
$144\mathrm{K}$ and $23\mathrm{K}$, respectively for CC-like and NC-like events. 
As described in details in the last published paper~\cite{chlast},
no $\tau^-$ decay candidates have been found so far.
The expected number of background events are
0.1 and 1.1 for the $\tau$ decay channels to $\mu^-$ and $h^-$, respectively.
The overall $90\%$ C.L.~upper limit on the number of $\tau$ decays is 2.4 and has been 
determined using the method proposed by
Junk~\cite{junk} which allows the combination of different channels, taking into account 
the errors on the background and on the signal.
This implies that oscillation parameters
lie outside the $90\%$ C.L.~curve shown in Figure \ref{exclus}, 
corresponding to an oscillation
probability $P_{\mathrm \mu\tau} < 3.4\times 10^{-4}$ and to $\stt < 6.8\times 10^{-4}$ for
large $\dmsq$.
\begin{figure}[t]
\begin{center}
\epsfxsize=28pc
\epsfbox{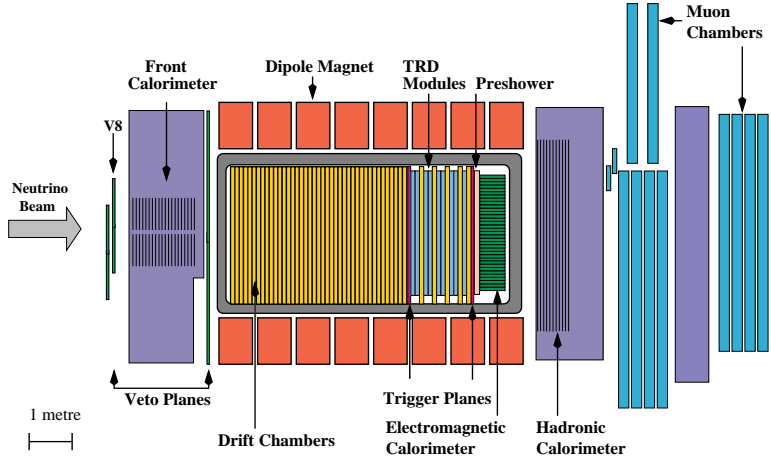}
\end{center}
\caption{\label{nomad1} Side view of the NOMAD detector.}
\vspace*{-3mm}
\end{figure} 

The CHORUS experiment is still analyzing neutrino interactions occurred 
into the emulsion target; at the 
end of the analysis it will reach a sensitivity corresponding,
in case of negative result and no background, to a limit on the 
$\numtonut$ oscillation probability of $P_{\mathrm \mu\tau} < 1.0\times 10^{-4}$.

Due to the electron-neutrino contamination in the beam, the absence of tau candidates
events can be used to quote also a limit on the $\nuetonut$ oscillation probability.
Using the above sample of analyzed events a limit of $P_{\mathrm e\tau} < 2.6\times 10^{-2}$
at $90\%$ C.L.~has been obtained. Full $\nuetonut$ mixing is excluded at $90\%$ C.L.~for
$\dmsqe>7.5$ eV$^2$; large $\dmsqe$ values are excluded at $90\%$ C.L.~for 
$\stte > 5.2\times 10^{-2}$.

\subsection{The NOMAD experiment}

The NOMAD detector is described in detail in reference~\cite{nomadet} and shown
in Figure~\ref{nomad1}. It consists of 
an active target made of drift chambers placed inside the former UA1 dipole magnet, with
a field of 0.4 Tesla perpendicular to the neutrino beam direction. The total mass
of the target is about 2.7 T. The target is followed by a transition radiation 
detector (TRD) and by a lead glass e.m. calorimeter, together giving a good
electron/hadron separation. 

The experiment took data from 1995 to 1998 collecting a sample of 
about $950\mathrm{K}\ \num$ neutrino charged-current interactions.

The NOMAD experiment searches for $\nut$ charged-current interactions by 
identifying $\tau^-$ decays
to $e^- \bar{\mathrm{\nu}}_e \nu_\tau$, inclusive decays to one or three charged 
hadron(s) + $\nu_\tau$ and exclusive decays to $\rho^-\nut$, for a total branching ratio
fraction of about $82 \%$. Since the identification of these reactions is achieved 
using only kinematic criteria, a good particle identification and a precise momentum 
measurement of all secondaries is required.

The criteria to select $\tau^-$ decays among charged and neutral current $\nu_\mu$
interaction relies on the presence of isolated $\tau$ decay products in the final 
state and on
the correlation in the transverse plane among their momentum, 
the total hadronic
system momentum and the missing momentum.
A special search for $\tau^-$ decays is also performed in a specific sample of events
characterized by low primary track multiplicity (less than 4),
enriched by quasi-elastic and resonance production.
\begin{figure}[t]
\begin{center}
\epsfxsize=24pc
\epsfbox{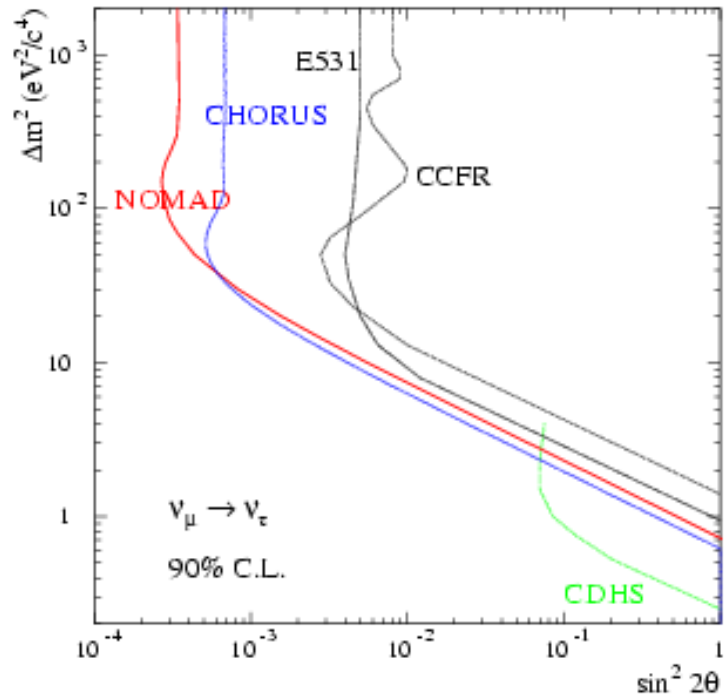}
\end{center}
\caption{\label{exclus} Chorus and NOMAD exclusion curves in the $\numtonut$ 
two flavor parameter space.}
\end{figure}

The combined analysis of all different $\tau$ decay modes shows no evidence for
$\numtonut$ oscillation~\cite{nomad}. 
To obtain its final result the NOMAD collaboration has adopted the technique proposed
by Feldman and Cousins~\cite{fc}.
The resulting $90\%$ C.L. upper limit on the oscillation probability 
is shown in Figure~\ref{exclus};
at present $P_{\mathrm \mu\tau} < 1.7\times 10^{-4}$, 
which corresponds to $\stt < 3.4\ \times 10^{-4}$ for large $\dmsq$. The
sensitivity of the experiment is $2.5\times 10^{-4}$; this is higher than the
quoted confidence limit, since the number of observed events is smaller than
the estimated background. In the absence of signal events, the probability
to obtain an upper limit of $1.7\times10^{-4}$ or lower is $39\%$.

A similar analysis has been performed in order to check the $\nuetonut$ hypothesis;
the $90\%$ C.L. upper limit on the oscillation probability is 
$P_{\mathrm e\tau} < 0.8\times 10^{-2}$, which corresponds to 
$\stte < 1.6\times 10^{-2}$ for large $\dmsqe$. The experiment sensitivity
is $1.2\times10^{-2}$.
In the absence of signal events, the probability
to obtain an upper limit of $0.8\times10^{-2}$ or lower is $43\%$.

\section{Conclusions}

The outcome of the CERN short baseline $\numtonut$ oscillation programme
was presented. Particle physics motivations based
on the see-saw mechanism and on cosmological arguments were at the basis
of CHORUS and NOMAD experiments, carried out at CERN to search mainly
for small mixing angles and relatively high mass differences 
in the $\mu - \tau$ sector. The results obtained so far by these two 
experiments do not show evidence of neutrino flavor oscillation.

\end{document}